\documentclass[conference]{IEEEtran}
\IEEEoverridecommandlockouts

\usepackage{cite}
\usepackage{amsmath,amssymb,amsfonts}
\usepackage{algorithmic}
\usepackage{graphicx}
\usepackage{textcomp}
\usepackage{xcolor}
\usepackage{subcaption}
\usepackage{url}
\usepackage{hhline}
\def\BibTeX{{\rm B\kern-.05em{\sc i\kern-.025em b}\kern-.08em
    T\kern-.1667em\lower.7ex\hbox{E}\kern-.125emX}}
\begin{document}

\title{CTA-Pipelining: A Latency-Oriented Spatial Scaling Method for Multi-GPU Systems}

\author{



\IEEEauthorblockN{
    Tingkai Liu\IEEEauthorrefmark{1}\IEEEauthorrefmark{2},
    Muralidhar Andoorveedu\IEEEauthorrefmark{2}, 
    Sanjoy Das\IEEEauthorrefmark{2},
    Sanjay Patel\IEEEauthorrefmark{2},
    Volodymyr Kindratenko\IEEEauthorrefmark{1}
}
\IEEEauthorblockA{
    \IEEEauthorrefmark{1} University of Illinois Urbana-Champaign
    \IEEEauthorrefmark{2} NVIDIA Corporation
    \\
    \{tingkai2, kindrtnk\}@illinois.edu,
    \{tingkail, mandoorveedu, sanjoyd, sanjpatel\}@nvidia.com
}
}

\maketitle

\begin{abstract}

The evolution of compute infrastructure has transformed multi-GPU systems into tightly integrated shared-memory structures. However, current software still mostly treats these coherent interconnects simply as high-speed networks. Simultaneously, the demand for serving Large Language Models under latency constraints has shifted GPU workload optimization from being throughput-driven to latency-bound, necessitating latency-oriented scaling methods beyond Tensor Parallelism (TP).

Thus, we introduce CTA-pipelining, an execution paradigm designed to exploit shared-memory multi-GPU systems. As a latency-oriented spatial scaling technique, CTA-pipelining leverages dependencies at the Cooperative Thread Array level, enabling concurrent execution of dependent kernels across GPUs. We demonstrate its capability using CUTLASS, cuBLAS, and NCCL libraries on 8-GPU H200 and B200 systems. Results show on 2-layer GEMM, representing the MLP operation, CTA-pipelining reduces latency by up to 31.8\% compared to micro-batching, and 29.6\% compared to TP. It can also be combined with TP as an orthogonal scaling dimension to further push the latency boundary.

\end{abstract}


\section{Introduction} 

Since the development of the Transformer architecture~\cite{vaswani2017attention, kaplan2020scaling}, optimizing Large Language Model (LLM) inference for production has become a critical challenge~\cite{2025taming}. Modern serving frameworks have two main objectives: maintaining high aggregate throughput for cost efficiency, and meeting latency Service-Level Objectives~\cite{agrawal2024sarathi}. In highly interactive scenarios, the limiting factor becomes single-batch user input latency. Since GPUs have traditionally been designed as throughput-oriented devices, minimizing latency for a single-batch request introduces new system-level requirements.

Correspondingly, to support the rapid scaling of LLM workloads, modern multi-GPU hardware systems have evolved into tightly coupled architectures, calling for novel software paradigm to fully exploit them. Systems such as the NVIDIA GB200 NVL72~\cite{nvidia2024gb200nvl72} utilize NVLink and NVSwitch~\cite{nvidia2024nvlink} interconnects not only to provide high peer-to-peer bandwidth, but also to enable the multi-GPU cluster to function with a unified shared-memory space. Although recent advancements have introduced sophisticated serving frameworks to maximize multi-GPU deployment efficiency~\cite{kwon2023vllm, zheng2023sglang, llmd2025, aibrix2025, nvidia2025dynamo}, as well as efforts to build systematic abstractions for programming multi-GPU workloads~\cite{sul2025parallelkittens, heldens2022lightning}, there remains untapped potential for novel execution paradigms that natively exploit these tightly coupled clusters as holistic shared-memory systems.

At the current stage, the standard paradigm for large-scale LLM deployment on multi-GPU systems relies on hybrid parallelism strategies~\cite{smith2022megatron, zheng2022alpa, narayanan2021efficient, qi2025synergistic}, primarily combining Pipeline Parallelism (PP)~\cite{huang2019gpipe, narayanan2019pipedream} and Tensor Parallelism (TP)~\cite{shoeybi2019megatron}. While emerging techniques such as Expert Parallelism (EP)~\cite{gale2023megablocks} and disaggregated serving~\cite{zhong2024distserve} offer further optimizations, they are highly workload-specific. Therefore, PP and TP remain the universal baselines. PP mainly focuses on improving overall serving throughput by operating at the inter-layer level to distribute transformer blocks across devices. TP provides both throughput improvement and latency reduction by spatially sharding computations at the operator level. However, TP introduces additional collective communication (e.g., AllReduce) to resolve data dependencies~\cite{zheng2022alpa}, establishing a hard ceiling on latency optimization.

Between inter-layer PP and operator-level TP lies an opportunity for further latency reduction: the \textit{intra-layer, inter-operator} execution space. Existing efforts accelerate this space primarily through temporal optimizations to improve efficiency on single device, such as kernel fusion~\cite{dao2022flashattention} or mega-kernels~\cite{aminabadi2022deepspeed, wu2024mirage,cheng2025mpk}. However, these require complex compiler toolchains or rigid rewrites~\cite{cheng2025mpk}. Alternatively, localized micro-batching~\cite{agrawal2023sarathi, zhong2024distserve} offers finer-grained pipelining, but introduces pipeline bubbles and degrades kernel efficiency at small chunk sizes. These limitations call for more efficient inter-operator spatial scaling techniques.

To address the dual requirements of single-batch latency optimization for GPU workloads and the need for new software paradigms for shared-memory multi-GPU systems, we propose CTA-pipelining, a novel latency-oriented spatial scaling method. Operating at the inter-operator level, it leverages the unified NVLink memory domain to enable simultaneous execution of data-dependent kernels across GPUs, via dynamic spatial pipelining at the Cooperative Thread Array (CTA) granularity. The designed protocol is minimally invasive to existing GPU kernel implementation, by relying only on the addition of prologue and epilogue code snippets. This preserves the potential for automated integration across a wide variety of workloads.


As an initial demonstration, we implement prototype and perform analysis using multi-layer general matrix-matrix multiplication (GEMM), a critical GPU workload, which also represents the multilayer perceptron (MLP) layers in transformer-based LLMs. Our implementation and evaluation are built upon the state-of-the-art NVIDIA libraries including CUTLASS~\cite{cutlass}, cuBLAS~\cite{cublas}, and NCCL~\cite{nccl}, on cutting-edge hardware including 8-GPU H200 NVLink system and 8-GPU B200 NVLink system.

We evaluate CTA-pipelining via two angles: validating the fundamental mechanism and demonstrating its broader scaling capabilities. First, to validate the protocol's overhead, we analyze the integration with both classical and warp-specialized multi-stage persistent CUTLASS kernels. Results show the overhead is minimal, and can even be mostly hidden within warp-specialized executions. Second, to establish CTA-pipelining as a generalized spatial scaling paradigm, we evaluate CTA-pipelining against traditional micro-batch chunk pipelining (achieving up to 31.8\% latency reduction) and Tensor Parallelism (achieving up to 29.6\% latency reduction), on setups representing MLP operations. We further demonstrate that it can be combined with TP as an orthogonal spatial scaling dimension, further pushing the latency scaling limit by providing benefit on both computation and communication.

The contributions of this paper are summarized as follows:
\begin{itemize}
    \item We propose CTA-pipelining, a novel spatial scaling paradigm for multi-GPU shared-memory systems, aimed at optimizing single-batch latency for multi-GPU workloads such as LLM inference.
    \item We demonstrate a prototype implementation of CTA-pipelining using the state-of-the-art GEMM library, NVIDIA CUTLASS, on various styles of GPU kernels.
    \item We perform analysis on basic protocol overhead, and identify the benefit of integrating with warp-specialized multi-stage persistent kernels.
    \item We compare CTA-pipelining against traditional micro-batch chunk pipelining, yielding latency reductions of up to 31.8\% on multi-layer GEMMs. 
    \item We compare CTA-pipelining against Tensor Parallelism on multilayer perceptron setup, providing latency reductions of up to 29.6\%. 
    \item We show how CTA-pipelining serves as an orthogonal spatial scaling method to Tensor Parallelism, offering benefit on both computation and communication, further pushing the latency optimization frontier. 
    \item We hint potential hardware evolutions that could further benefit this execution model.
\end{itemize}


\section{Background and Related Work} 

\subsection{GPU Execution model}

Basic understanding of the GPU execution model is essential to understand this work. CUDA~\cite{cuda} uses Single Instruction, Multiple Data (SIMD) execution model~\cite{lindholm2008nvidia}. Each GPU kernel launches a global grid of threads to process data concurrently. These threads are grouped into blocks, also known as Cooperative Thread Arrays (CTAs). Each CTA utilizes a localized shared memory to cooperatively compute a specific tile of data. Recent architectures further introduce a abstraction layer called Thread Block Clusters (or Cooperative Group Arrays, CGAs), which groups multiple CTAs together, enabling cooperate via Distributed Shared Memory (DSMEM)~\cite{nvidia2022hopper}. At the hardware level, CTAs are dispatched to Streaming Multiprocessors (SMs), where threads are bundled into groups of 32, known as warps. While individual warps are scheduled and operate independently, threads within the same warp execute the same instruction simultaneously.

As GPU architectures evolve, core computational workloads are increasingly offloaded to specialized on-chip accelerators. General-purpose SMs are transitioning from performing the primary computation to orchestrating these specialized units. Mature accelerators include Tensor Core for Matrix-Matrix Multiply-Accumulate (MMA), and Tensor Memory Accelerator (TMA). The recently announced integration with Language Processing Unit (LPU) for LLMs also reflects this trend~\cite{LPU}. To efficiently exploit this heterogeneous on-chip setup, warp specialization has emerged as a standard programming paradigm. Specific warps are dedicated to each stages of an execution pipeline, managing asynchronous calls to the hardware accelerators. High-performance GEMM implementations, such as those in the NVIDIA CUTLASS library~\cite{cutlass}, rely on this paradigm to optimize the overall computation.


\subsection{Inter-operator Optimizations}

Large amount of existing work focuses on improving GPU execution efficiency through the co-optimization of consecutive kernels (inter-operator optimization). The boundaries between discrete kernels often introduce performance bottlenecks, including kernel launch overheads~\cite{shen2020nimble}, redundant global memory round-trips~\cite{le2023welder}, and wave quantization effects, where the final wave of scheduled thread blocks fails to fully utilize the available resource. Furthermore, individual kernels have different resource utilization profiles, typically categorized as compute-bound, memory-bound, or communication-bound. By fusing or concurrently executing these operators, it is possible to overlap mismatched resource demands, achieving higher overall hardware utilization.

To directly address inter-operator inefficiencies, kernel fusion approaches this by systematically merging multiple consecutive operations into a single kernel~\cite{chen2018tvm, li2022automatic}. Pushing this optimization paradigm to its extreme results in the mega-kernel approach. In a mega-kernel design, massive computational sub-graphs, ranging from complex attention mechanisms~\cite{dao2022flashattention} to entire multi-GPU model inference pipelines~\cite{wu2024mirage,cheng2025mpk}, are encapsulated within a single GPU kernel launch. However, this extreme form of fusion requires re-compilation of the whole workflow, as well as sophisticated in-kernel runtimes to manage intra-kernel synchronization, warp-level task scheduling, and decentralized hardware resource allocation to prevent SM underutilization~\cite{ma2020rammer}. 

Besides, inter-operator pipelining offers a complementary approach by overlapping consecutive execution stages. Unlike coarse-grained inter-layer parallelism, these techniques decompose individual operators into fine-grained micro-batches or tiles to enable concurrent processing~\cite{zheng2017versapipe,  cusync2024, zhang2025hytis}. This intra-layer pipelining also serves as a primary mechanism for achieving communication-computation overlap~\cite{zhu2025nanoflow, chang2024flux, wang2023overlap}. 

However, these methods have primarily been deployed as temporal pipelining on single devices to improve overall hardware utilization. The performance benefits of some of those approaches also rely on processing multiple batches of data concurrently, for example, by overlapping the communication of a previous batch with the computation of the current one. The use of fine-grained inter-operator pipelining spatially across multiple devices for reducing single-batch input latency remains largely unexplored. Historically, this is mainly because pipelining introduces pipeline bubbles and requires frequent inter-stage synchronizations, limiting its effectiveness on latency reduction~\cite{zheng2022alpa}.

Very recently, hardware-centric studies such as Kitsune~\cite{davies2025kitsune} have begun exploring the hardware architecture potential to support spatial pipelines. Our work comes from a different angle: we propose a pure software technique designed to exploit the capabilities of modern shared-memory multi-GPU systems, focusing on using fine-grain spatial pipelines to reducing single-batch query latency.

\section{CTA-Pipelining Protocol: Design, Integration with CUTLASS, and Overhead Analysis}

In this section, we present the design and implementation of the CTA-pipelining protocol. We demonstrate its integration into the state-of-the-art NVIDIA CUTLASS library for GEMM operations and provide a detail execution trace analysis to understand and quantify the protocol's overhead.

As a software approach built upon current hardware and CUDA programming capabilities, our inter-kernel communication relies on atomic counters and queues, sharing structural similarities with latest literatures~\cite{cusync2024,davies2025kitsune}. However, we distinguish our work by elevating this mechanism into a cross-device spatial scaling method, and provides the placement strategy and additional coherency-ensuring design for multi-GPU setup. Moreover, we present novel implementation of integrating the protocol into multi-stage warp-specialized persistent kernels, and provide novel finding that such integration can effectively hide protocol overhead. 

\subsection{Basic Protocol}




As its name suggests, CTA-pipelining is designed to execute dependent GPU kernels in a deeply pipelined manner at the finest architectural granularity: the CTA. In this model, each CTA consumes one or more input data tiles and produces a single output tile. Rather than enforcing strict kernel-level synchronization, our approach allows the consumer kernel to launch its CTAs as soon as the producer kernel generates the requisite data tiles. The kernel execution trace is similar to Figure~\ref{fig:chunk-cta}, where data dependent kernels are spatially distributed across GPU compute resources, getting launched simultaneously, and complete almost at the same time, except the tail at the consumer kernel for the last wave of pipelined CTA computation. 

In single-GPU context, this goal is conceptually similar to the Megakernel~\cite{wu2024mirage,cheng2025mpk} approach, which recompiles the entire workload to execute a tile-level dependency graph on a single device. However, CTA-pipelining aims to enable fine-grained pipelined execution spatially across multiple GPUs while preserving the original kernel structure, without the need to recompile the overall workflow. To enable data-dependent kernels to launch concurrently across multiple GPUs while maintaining execution correctness, additional control-flow dependency organization is required.

\begin{figure}[t]
    \centering
    \includegraphics[width=\columnwidth]{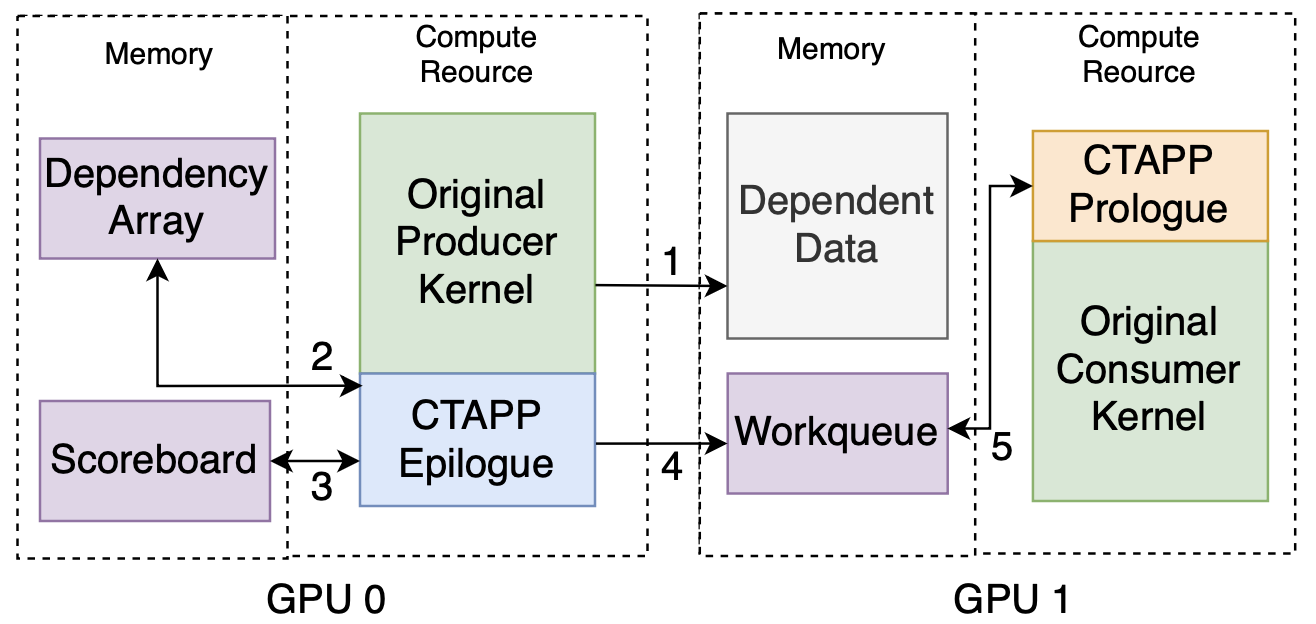}
    \caption{Setups of enabling CTA-pipelining across 2 GPUs.}
    \label{fig:ctapp-setup}
    \vspace{-3mm}

\end{figure}

Figure~\ref{fig:ctapp-setup} illustrates the overall execution process of CTA-pipelining, using a two-GPU setup to demonstrate how the original kernel interacts with the additional components. The underlying data structures that support this paradigm include dependency arrays, scoreboards, and inter-device workqueues, which we collectively refer to as \textit{dependency structure}. To interface with the dependency structure and orchestrate the control flow, lightweight prologues and epilogues are injected directly into the original kernel code. We detail each of these core components below, followed by a step-by-step execution walkthrough.

\subsubsection{Dependency Array}
The dependency array indicates the control-flow dependencies between producer CTAs and consumer CTAs. Specifically, it is indexed by the producer CTA ID to identify which consumer CTAs if affects, including a contiguous list of consumer CTA IDs and a corresponding offset array that defines the index range for each producer. These dependencies can be derived directly from data dependencies, using static analysis, trial runs of the kernels, or even calculated dynamically at runtime if closed-form formula exist. As the dependency array is exclusively accessed by consumer CTAs, it is stored on the producer device memory.

\subsubsection{Scoreboard}
To resolve many-to-one data dependencies, where a consumer CTA relies on output from multiple producer CTAs, a scoreboard is used to track producer completion. The scoreboard consists of an array of atomic counters, with each consumer CTA assigned an entry,  initialized to the total number of its prerequisite producer CTAs. A counter reaching zero serves as a readiness signal, indicating a consumer CTA is ready for execution. The scoreboard is initialized as part of the prior dependency analysis. Since the scoreboard is also exclusively modified by producer CTAs, it is stored on the producer device memory.

\subsubsection{Inter-device Workqueue}
The inter-device workqueue serves as a readiness signaling mechanism between the producer kernel and the consumer kernel. Implemented as a cyclic ring buffer, it is managed by several atomic values including head, tail, and size to ensure coherency. 

Since the workqueue requires concurrent access from both kernels, cross-device NVLink is utilized. We place the workqueue in the consumer device's memory to minimize critical-path latency. While cross-device writes are theoretically expensive, CUDA's asynchronous write semantics allow the producer to issue "fire-and-forget" memory operations without stalling for completion (unless there is explicit flush barrier). Conversely, consumer CTAs must continuously poll the workqueue, and forcing these frequent read operations across NVLink would incur latency penalties. Therefore, placing the workqueue on the consumer device is a more reasonable design choice.

\subsubsection{Overall Workflow with Prologue and Epilogue Code Snippets}

The data structures described above facilitate the CTA-pipelining control flow. The logic that utilizes these data structures consists of minimal prologue and epilogue code snippets. These are added to the beginning and end of the kernel code, leaving the core original kernel implementation untouched. (Warp-specialized kernels are treated slightly differently, as introduced later.)

The overall control flow orchestrated by the injected prologue and epilogue is illustrated by the numbered arrows in Figure~\ref{fig:ctapp-setup}. First, the producer CTA writes its output directly to the consumer's input memory across NVLink, as indicated by arrow (1). Its epilogue then issues a system-wide memory fence, ensuring subsequent dependency operations and workqueue updates do not become visible to other devices before the actual output data is visible. After that, it utilizes SIMD execution to atomically decrement dependency counters in scoreboard after querying dependency array (2, 3). When a counter reaches zero, the producer thread pushes the ready consumer CTA ID into workqueue (4). On the consumer side, the injected prologue uses a single thread to busy-poll this workqueue while the remaining threads wait at a barrier (5). Upon retrieving a ready ID, the polling thread broadcasts it via shared memory. The consumer CTA then remaps its ID to this fetched value and executes its standard, unmodified kernel payload.

\subsubsection{Host-Side Organization}

From the host side, each kernel is assigned a dedicated CUDA stream and bound to a specific device or a partition of devices. During execution, all kernels are launched simultaneously, with internal execution order dynamically guided by the CTA-pipelining protocol. In multi-layer workloads, intermediate kernels act as both producers and consumers. These kernels feature both a prologue to fetch from an source workqueue and an epilogue to push to a destination workqueue. Note, this entire multi-kernel execution process can be captured by CUDA Graphs, reducing kernel launching bubbles. 

In actual workflows, the CUDA driver's default CTA scheduling order might not allow the entire workload to execute in a smoothly pipelined manner. In such cases, altering the execution order, such as shifting from a column-major to a row-major pattern, can be beneficial. In this case, the first kernel in the workflow can simply consume a pre-defined source workqueue to explicitly guide its CTA execution order.

There is an optional micro-optimization to eliminate the SM resource waste of initial busy-polling, using \textit{cuStreamWaitValue32} API. This feature blocks the consumer's execution stream until the workqueue contains at least one item, preventing immediate spin-waiting. However, this stream-level synchronization does incur a slight latency penalty for the kernel launch.


\subsection{Integration with CUTLASS GEMM Kernels}

In this subsection, we explain how the CTA-pipelining protocol can be integrated into two different styles of GPU kernels, using NVIDIA CUTLASS GEMM implementations~\cite{cutlass} as examples. To illustrate the dependency mapping, we use a two-layer GEMM workload as our running example, where the output of the first GEMM is directly consumed by the subsequent GEMM. 

\subsubsection{Classical Kernels}

As a straightforward example, we demonstrate how CTA-pipelining integrates with classical GEMM kernels, using the SM90 TMA kernel from the CUTLASS library. In this kernel, each CTA has all threads follow the same execution flow, producing one tile of the output matrix. At the end of its execution, the CTA terminates, and the next wave of CTAs is scheduled by the CUDA driver. This represents the most classical GPU kernel structure.


In terms of control dependencies, each consumer CTA computes one output tile by consuming an entire row of input tiles, making it dependent on an entire row of producer CTAs. Furthermore, consumer CTAs in the same row share identical dependencies. While we recognize that this could be simplified to a row-to-row dependency rather than a strict CTA-to-CTA dependency, or calculated dynamically at runtime using matrix and tile dimensions, we present the most general case for illustrative purposes.

Regarding modifications to the CUTLASS library, integrating CTA-pipelining protocol is minimally invasive. Beyond extending the parameter structure to accept the necessary dependency structures, the source code changes are restricted to injecting the prologue and epilogue snippets. Notably, while the prologue requires shared memory to broadcast the dynamically fetched CTA ID to all threads, it reuses the memory space already allocated for the kernel's main execution. This does not affect the subsequent execution, as the prologue runs at the very beginning of the kernel and its shared memory usage is strictly one-time.

In summary, integrating CTA-pipelining with classical kernels is highly straightforward. It does not require a detailed understanding or modification of the core execution logic, demonstrating that this approach is highly generalizable to other kernels.

\subsubsection{Warp-Specialized Multi-Stage Persistent Kernels}

The warp-specialized multi-stage persistent kernel is a representative example of programming models evolution and serves as the foundational design for modern high-performance libraries like CUTLASS. As its structural paradigm differs from classical kernels, it is helpful to demonstrate how CTA-pipelining integrates with this kernel design, using SM100 TMA Warp Specialized kernel from CUTLASS as an example. 

This architecture relies on three important concepts. First, \textit{persistent kernels} launch a fixed number of CTAs that remain active for the entire computation, dynamically fetching work-tiles via explicit metadata rather than relying on standard CTA IDs. Second, a \textit{multi-stage} design divides the workflow into an internal micro-pipeline. Third, \textit{warp specialization} assigns these distinct micro-pipeline stages to independent warps of threads.


In the SM100 CUTLASS implementation, these three concepts operate together. The resulting micro-pipeline includes discrete stages for scheduling, main data loading, Matrix Multiply-Accumulate (MMA), GEMM epilogue data loading, and GEMM epilogue operations. One dedicated warp handles each stage, except the GEMM epilogue stage might have multiple warps.  

The CTA-pipelining prologue is integrated directly into the scheduler warp. In the original SM100 CUTLASS kernel, information regarding the next work tile is obtained via a hardware query, known as a Cluster Launch Control (CLC) query, issued by a single thread within the scheduler warp to fetch the next ready work ID. Subsequently, the returned information is multicast across the CGA and stored in each CTA's Distributed Shared Memory (DSMEM). After this broadcast, each active warp decodes the response to extract the index of its next work tile.

Aligning with this native kernel structure, the CTA-pipelining prologue reuse this work-tile fetching and broadcasting pathway. Instead of issuing a hardware CLC query, the scheduling thread performs busy-polling on the added workqueue. Once it successfully fetches a work item from the queue, it issues remote memory stores to the DSMEM across the CGA, reusing the buffer originally allocated for the CLC response. Since the scheduler tracks the total number of tiles to be executed, it uses this information to determine the completion of the workload by comparing the current workqueue index against the total tile count. The other compute warps fetch the next work tile similar to before, but changing the decoding format for the new metadata structure.

The CTA-pipelining epilogue is positioned at the final stage of the micro-pipeline: the GEMM epilogue. In the SM100 CUTLASS GEMM implementation, multiple warps are used to execute arithmetic operations, while only one warp performs the final output write to global memory. As the global threadfence that ensures consistent memory views is only effective for each calling thread, only the specific warp responsible for writing global output issues the threadfence, and only this warp is utilized for the CTA-pipelining epilogue operations. The specific operations performed on the dependency structure remain identical to before.

A special case arises with Blackwell's specific Tensor Core operations, where two CTAs cooperatively issue a single Tensor Core instruction. In the CUTLASS GEMM implementation, this is reflected as a change in the tile ID mapping: while each CTA still receives separate work-tile information, the total tile size is doubled along one dimension to accommodate the CTA-pair operation. To handle this special case, the work-tile information fetched from the workqueue are decoded differently, and the scoreboard tracking values are adjusted accordingly. All other parts of the pipelining logic remain unchanged.

\subsection{Protocol Overhead on CUTLASS}

The first question that naturally arises is how the additional operations introduced by the CTA-pipelining protocol affect overall performance. To understand this impact, we investigate detailed execution phases and profile the latency of each operation by injecting timestamping code into the kernel, and perform analysis based on the collected timestamps. We present results for both classical kernels and warp-specialized multi-stage persistent kernels, as they exhibit different performance patterns. 

The evaluation of the classical SM90 CUTLASS kernel is performed on a 8-GPU H200 system, with GPUs connected by the $4^{th}$ generation NVLink. The SM100 TMA Warp Specialized kernel is evaluated on a 8-GPU B200 system, where GPUs are connected by the $5^{th}$ generation NVLink. For all evaluation on protocol overhead, all GEMM input sizes are 16384 by 8192, multiplying with weight matrix of 8192 by 8192. The input and output data types are set to BF16, and the Tensor Core accumulator type is set to FP32. The specific CUTLASS configuration is selected by running the CUTLASS Profiler~\cite{cutlass} on the given input size, and selecting the configuration with the fastest execution time.  

\begin{figure}[b]
    \centering
    
    \begin{subfigure}[b]{0.9\columnwidth}
        \centering
        \includegraphics[width=\columnwidth]{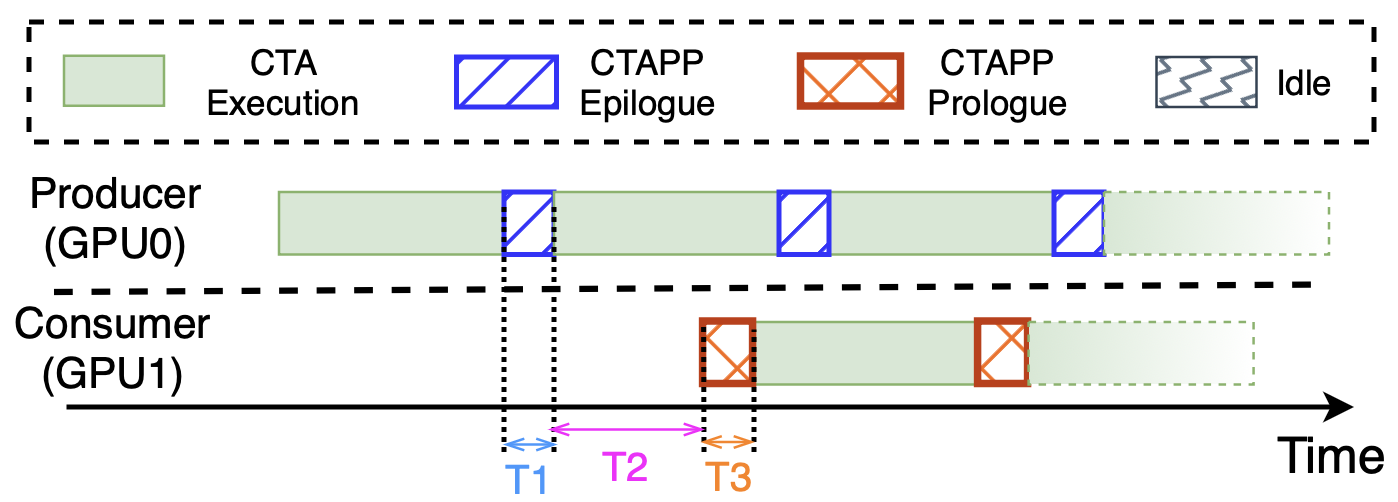}
        \caption{}
        \label{fig:pic-basic}
    \end{subfigure}

    \begin{subfigure}[b]{0.9\columnwidth}
        \centering
        \includegraphics[width=\columnwidth]{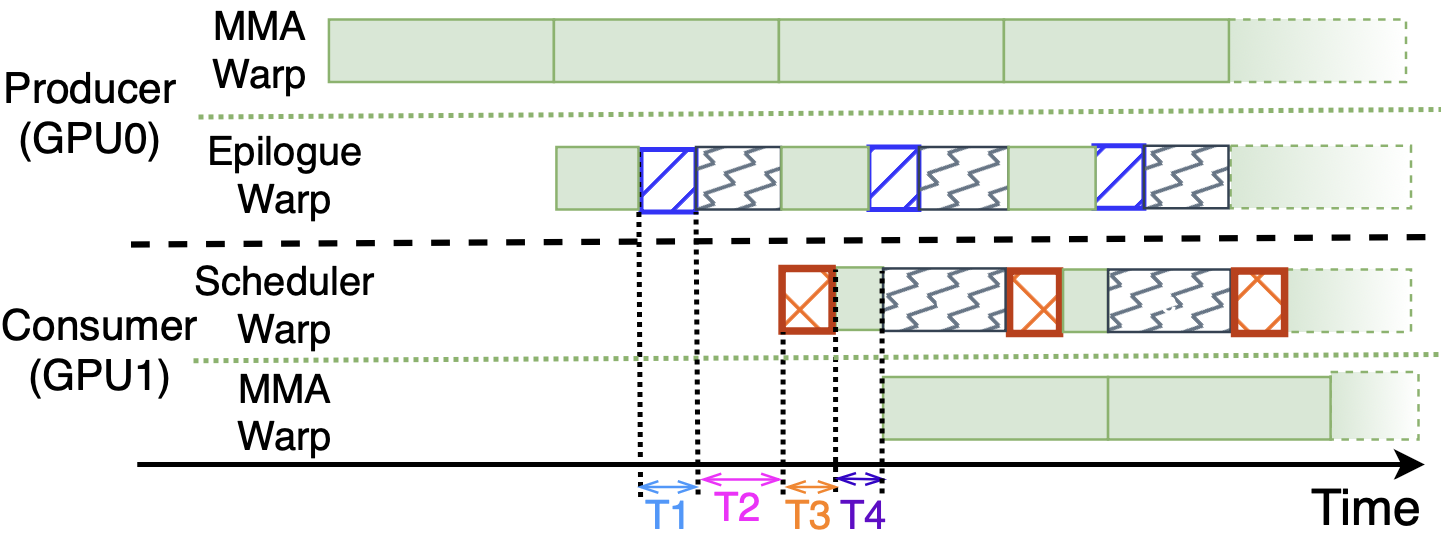}
        \caption{}
        \label{fig:pic-warp}
    \end{subfigure}
    
    \caption{The detail execution trace of adding CTA-pipelining to (a) classical kernel, and (b) warp-specialized multi-stage persistent kernel. For illustration purpose, time is not at the correct scale.}
    \label{fig:pic}

    \vspace{-3mm}

\end{figure}

For the classical kernel, the profile of the execution phases is shown in Figure~\ref{fig:pic-basic}. For each CTA, the epilogue operations take about T1=6$\mu$s. Subsequently, it takes T2=120$\mu$s for this update to become visible to the consumer device. On the consumer side, the prologue takes about T3=1.5$\mu$s to fetch the work queue value and broadcast the fetched information. Notice, this overhead accumulates for each wave of CTA execution.

The warp-specialized persistent kernel exhibits dramatically different results. CTA-pipelining can utilize the idle time of specific warps, effectively hiding its protocol overhead. As introduced earlier, the warp-specialized kernel inherently functions as a micro-pipeline. Since the entire pipeline is typically dominated by the latency of the compute-heavy MMA operations, the other warps frequently need to wait at pipeline barriers. This presents an opportunity to perform additional work within these waiting warps, such as the CTA-pipelining operations, without affecting overall performance. 

The exact execution trace for the warp-specialized persistent kernel is illustrated in Figure~\ref{fig:pic-warp}, including only the related warps for analysis. For each wave of execution, the epilogue operations, including the threadfence, take approximately T1=6$\mu$s. During this time, other warps remain active for the next wave of computation. By the time the next computational wave reaches the GEMM epilogue warp, the CTA-pipelining epilogue operations from the previous wave have already finished, allowing the warp to take on the next wave immediately, which hides the CTA-pipelining operation overhead. The cross-device NVLink data write takes roughly T2=5$\mu$s to become visible on the consumer device. On the consumer side, the prologue work queue fetching takes about T3=1.5$\mu$s, and it takes about T4=0.5$\mu$s to broadcast across the CGA by the scheduler warp. Similarly, since the scheduler warp is an stage of the micro-pipeline, this latency is also hidden. As a result, assuming no other memory or communication contention, the CTA-pipelining overhead is theoretically only visible once during the initial pipeline ramp-up.

As the CTA-pipelining logic is confined to the injected prologue and epilogue, it also imposes minimal performance interference on the original kernel. It does not disrupt the main compute phase's active register allocation or shared memory usage, avoiding register spilling or shared memory contention. For instance, in our evaluation of SM100 TMA Warp Specialized kernel, the baseline execution of two consecutive GEMMs takes an average of 1080$\mu$s each. When executed using CTA-pipelining, the producer completes in 1090$\mu$s, and the consumer finishes in 1165$\mu$s, counting in both the visible overhead and the latency of the final pipeline wave.

In summary, the overhead introduced by the CTA-pipelining protocol is minimal, and it can even be largely hidden within warp-specialized persistent kernels, making it suitable for general fine-grain spatial pipelining. We demonstrate more scaling examples with performance analysis in the next section. 

\section{CTA-Pipelining as a Scaling Method: Comparing against Micro-batching and TP} 

In this section, we demonstrate how CTA-pipelining can be used as a latency-oriented spatial scaling method, on optimizing single batch execution latency. The evaluation consists of two parts. First, we perform a comparison against traditional static micro-batch chunk pipelining (also known as micro-batching), using a multi-layer GEMM as the primary workload. Second, we compare CTA-pipelining against Tensor Parallelism (TP), and demonstrate how it can be combined with TP by evaluating performance across different scaling setups and input sizes.

The workload chosen for this section consists of multi-layer GEMM operations, which simulate LLM MLP layers by omitting intermediate element-wise non-linear activations for simplicity. This abstraction assigns specific meanings to our matrix dimensions. For example, in the two-layer GEMM $Y=XAB$, $X$ represents the input activation tensor, while $A$ and $B$ are the respective weight matrices. Because of this mapping, we fix the weight matrices $A$ and $B$ at 8192 by 8192 to reflect typical LLM architectures, while leaving the row dimension of $X$ flexible for varying input sequence lengths. As before, the input and output data types are set to BF16, and the Tensor Core accumulator type is set to FP32, with specific CUTLASS configuration determined by the CUTLASS profiler. 

Our testbeds is an 8-GPU NVIDIA B200 system with $5^{th}$ generation NVLink. All the CTA-pipelining implementation utilize the integration with SM100 TMA Warp Specialized kernel introduced in previous section. The baseline we compare against are built using state-of-the-art cuBLAS~\cite{cublas} and NCCL~\cite{nccl} libraries. All execution time we present are the average of 5 repeated runs, where data are collected using NVIDIA Nsight Systems~\cite{nsight}. 

\subsection{Comparison Against Micro-Batch Chunk Pipelining}

Intuitively, the CTA-pipelining execution model is similar to micro-batching operating at the finest granularity. However, they are architecturally different, especially in terms of kernel efficiency and parallelism. Unlike traditional methods, CTA-pipelining does not explicitly pre-partition the input into static chunks, nor does it invoke multiple separate kernel launches for each stage of the pipeline. Instead, it preserves the original kernel structure and leverages the unified memory space within the NVLink domain for coordination. This approach preserves single-kernel efficiency while enabling cross-device parallelism, qualifying it as a general, latency-oriented spatial scaling method. In this subsection, we present experimental results that support these claims.

To compare against static micro-batching, we performed experiments running a multi-layer GEMM workload across multiple GPUs using our prototype built with NVIDIA CUTLASS. As a baseline, static micro-batching is built with the NVIDIA cuBLAS library. For this specific scenario, cuBLAS is considered a stronger baseline than CUTLASS, as it self-adapts to different input sizes, providing the most robust comparison when varying static chunk sizes. Furthermore, the baseline is executed using CUDA Graph to minimize kernel launching overhead. 

\begin{figure}[htbp]
    \centering
    
    \begin{subfigure}[b]{0.9\columnwidth}
        \centering
        \includegraphics[width=\columnwidth]{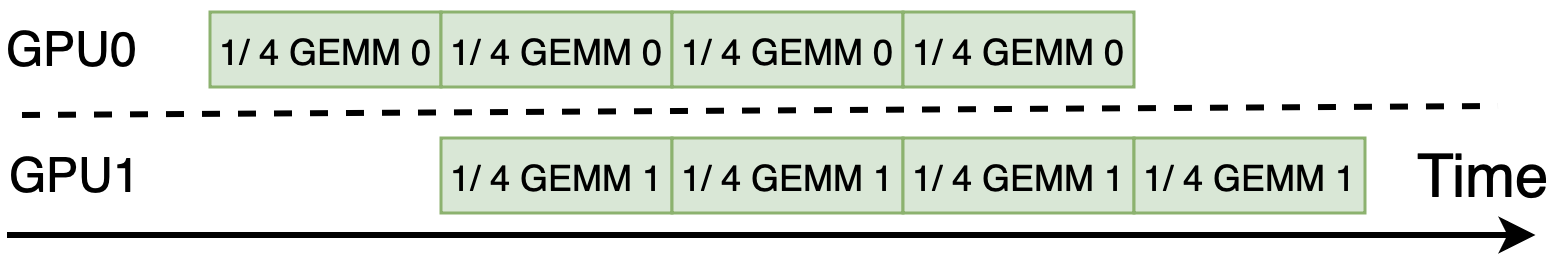}
        \caption{}
        \label{fig:chunk-basic}
    \end{subfigure}

    \begin{subfigure}[b]{0.9\columnwidth}
        \centering
        \includegraphics[width=\columnwidth]{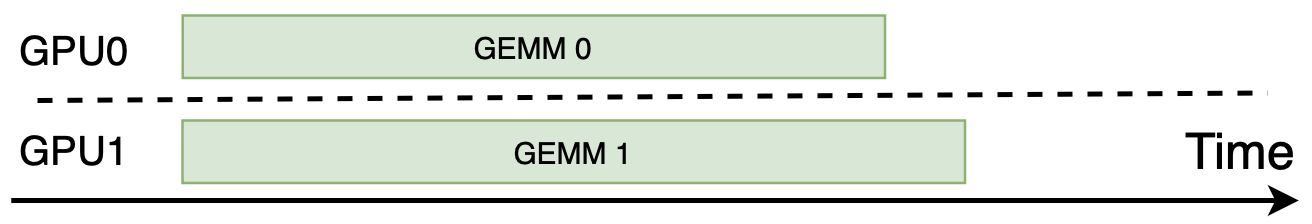}
        \caption{}
        \label{fig:chunk-cta}
    \end{subfigure}
    
    \caption{The experiment setup for comparing executing multiple layers of GEMM with (a) micro-batch chunk pipelining, and (b) CTA-pipelining, using GPU=2 as an example.}
    \label{fig:chunk-experiment}

\end{figure}

For demonstration, the input sequence length is chosen to be 16384, while the weight dimensions are fixed to be 8192 by 8192, as introduced before. Each GPU is assigned one layer of the GEMM workload, and increasing number of GPUs also increases the layer of GEMM operations. Micro-batching is performed by splitting the input matrix in row-major order, where the "chunk size" we refer to is the input sequence length (row dimension) of the split sub-matrix, and the number of pipeline stages is equal to the number of GEMM layers. This experimental setup is illustrated in Figure~\ref{fig:chunk-experiment}. 

\begin{figure}[t]
    
    \begin{subfigure}[b]{\columnwidth}
        \centering
        \includegraphics[width=\columnwidth]{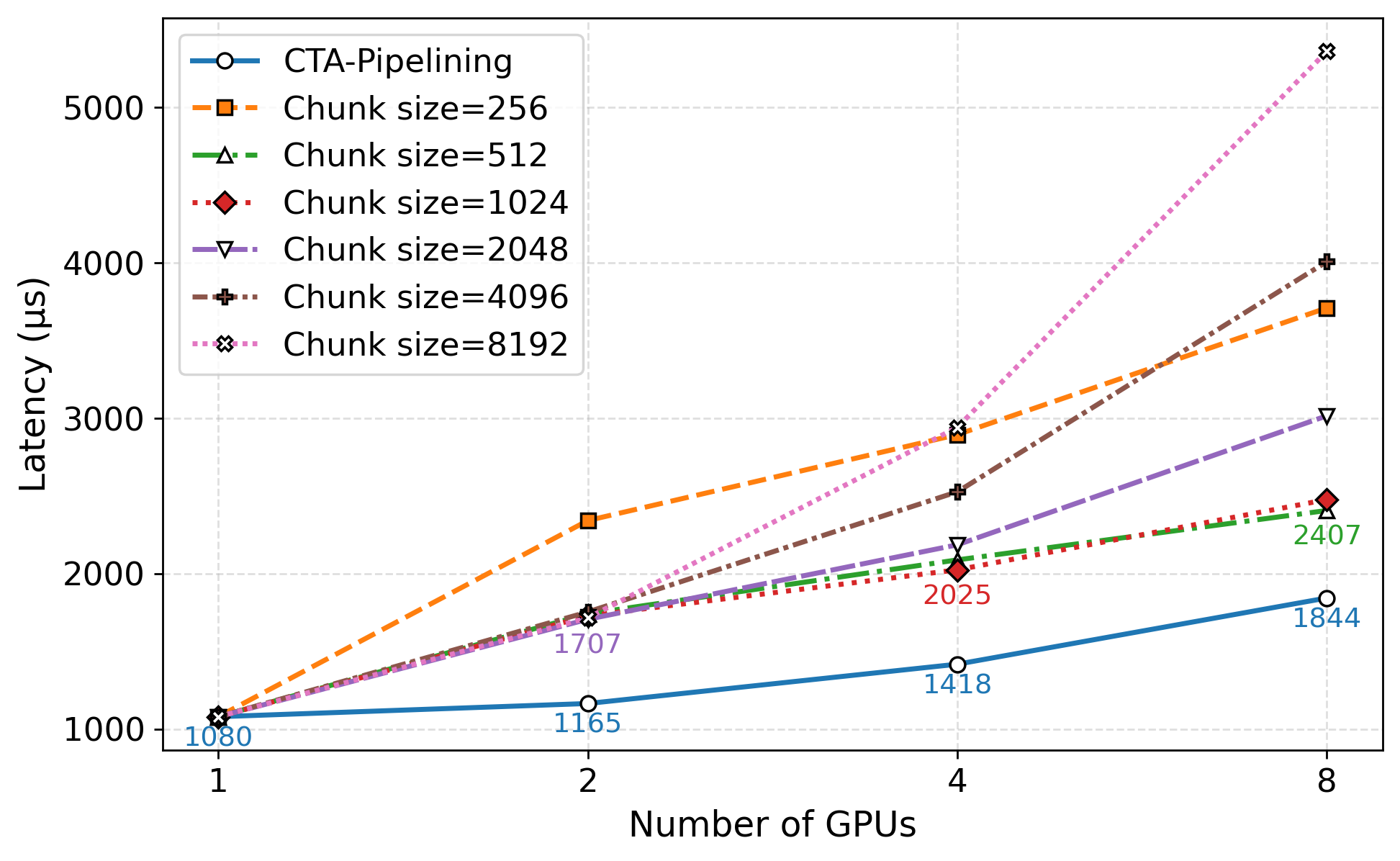}
    \end{subfigure}
    
    \begin{subfigure}[b]{\columnwidth}
        \centering
        \small
        \begin{tabular}{|c|c|c|c|c|}
            \hline
            Number of GPUs & 1 & 2 & 4 & 8 \\
            \hline
            Latency Reduction & - & 31.8\% & 30.0\% & 23.4\% \\ 
            \hline
        \end{tabular}
    \end{subfigure}
    
    \caption{Multi-layer GEMM (16384$\times$8192$\times$8192) latency reduction, comparing CTA-pipelining against micro-batching implemented with cuBLAS. The number of GEMM layers increases with the number of GPUs.}
    \label{fig:chunk-speedup}

    \vspace{-3mm}

\end{figure}

\begin{figure}[t]

    \begin{subfigure}[b]{\columnwidth}
        \centering
        \includegraphics[width=\columnwidth]{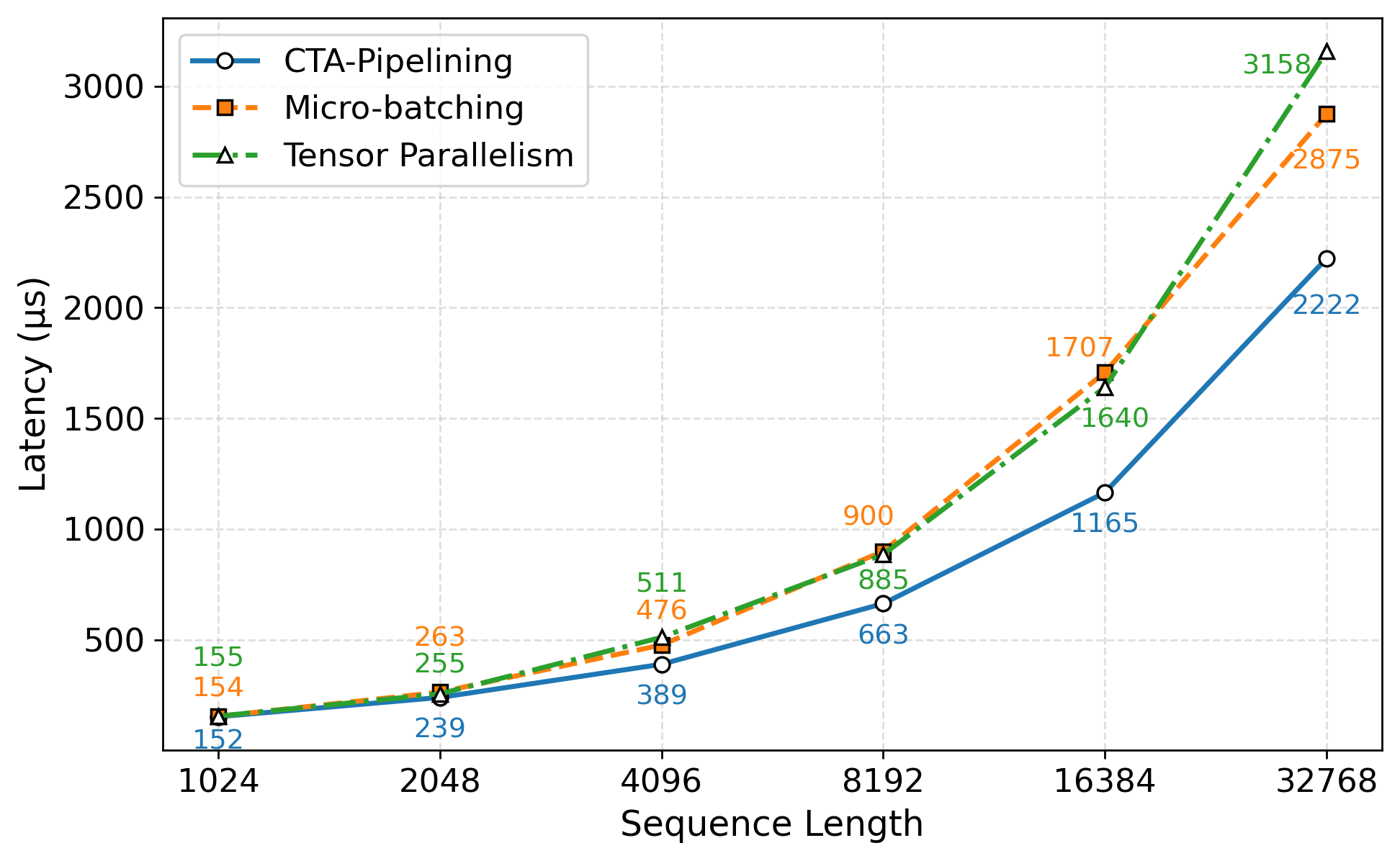}
    \end{subfigure}
    
    \begin{subfigure}[b]{\columnwidth}
        \centering
        \small
        
        \begin{tabular}{|c|c|c|c|c|c|c|}
            \hline
            \textbf{Input} & \textbf{1024} & \textbf{2048} & \textbf{4096} & \textbf{8192} & \textbf{16384} & \textbf{32768} \\
            \hhline{|=======|}
            R-MB & 1.3\% & 9.1\% & 18.3\% & 26.3\% & 31.8\% & 22.7\% \\ 
            \hline
            R-TP & 1.9\% & 6.3\% & 23.9\% & 25.1\% & 29.0\% & 29.6\% \\ 
            \hline
        \end{tabular}

        \flushleft
        R-MB: Latency reduction against micro-batching \\
        R-TP: Latency reduction against Tensor Parallelism \\
        $\ast$ The weight matrices are fixed to be 8192$\times$8192. \\
        $\ast\ast$ The micro-batching execution latency is obtained by sweeping chunk sizes and selecting the optimal one.
    \end{subfigure}

    \caption{CTA-Pipelining latency vs. micro-batching and Tensor Parallelism on 2-layer GEMM, representing MLP layers, on 2 GPUs, across different input sequence length.}
    \label{fig:2-layer}
    \vspace{-3mm}
    
\end{figure}

The experimental results are presented in Figure~\ref{fig:chunk-speedup}. We compare CTA-pipelining against static micro-batching across varying numbers of GPUs and a range of chunk sizes. As optimal chunk size selection is critical for the baseline's performance, sweeping this parameter ensures a fair comparison. The results demonstrate that CTA-pipelining consistently outperforms static micro-batching across all evaluated configurations. Comparing against the optimal chunk size in the sweeping, CTA-pipelining reduces latency by 31.8\%, 30.0\%, and 23.4\% in 2, 4, and 8 GPU setup, respectively. The more common 2-layer GEMM case, representing MLP layers, is further evaluated across different input sequence length in Figure~\ref{fig:2-layer}, further proving CTA-pipelining's effectiveness.

The fundamental problem with traditional micro-batch chunk pipelining can be viewed as a dilemma between parallelism and kernel efficiency when selecting chunk sizes. When the chunk size is too large, the head and tail phases of the pipeline execution become longer, where most devices waiting for the previous pipeline stage to finish, thereby decreasing overall parallelism. If the chunk size is too small, the compute efficiency of the individual kernels suffers. GPU kernels require sufficiently large input sizes to maintain high hardware utilization, and small kernels often suffer from quantization effects. Smaller chunk sizes also implies larger number of kernel launches, accumulating more kernel launching bubbles. 

In contrast, CTA-pipelining achieves overlap at the finest possible granularity: the CTA level. At the same time, as it does not explicitly split the input into discrete chunks, the original compute efficiency of the kernel is preserved. This effectively solves the traditional chunking dilemma. 

Furthermore, executing micro-batching as a spatially across devices introduces additional overhead. It suffers not only from pipeline bubbles induced by repeated kernel launches, but also from cross-device write latency, as the final wave of TMA writes must post before the kernel can terminate. In contrast, our execution trace analysis in Figure~\ref{fig:pic-warp} demonstrates that, while CTA-pipelining requires a system-wide memory fence to guarantee inter-device memory consistency, the warp-specialized kernel design effectively hides this overhead. 

Despite its significant benefits, CTA-pipelining does become less effective at extremely small input sizes. As shown in Figure~\ref{fig:2-layer}, while protocol overhead is mostly hidden, it does stand out when kernel execution time drops to the 100$\mu$s range (e.g., sequence length 1024). In the most extreme case, where a kernel requiring only a single wave of CTAs, intra-kernel pipelining becomes physically impossible. Nevertheless, traditional micro-batching is also reaching its limit under such extreme scenarios.

In summary, CTA-pipelining is a superior method than micro-batch chunk pipelining in most of the cases. It maximizes cross-device parallelism and preserves native kernel efficiency, while effectively hiding pipeline bubbles and global memory fence overheads. As an added practical benefit, the CTA-pipelining approach saves the tuning effort typically required to discover the optimal micro-batch chunk size. For workflows that already utilize micro-batch pipelining executing, CTA-pipelining can serve as a direct replacement.

\subsection{Comparison Against Tensor Parallelism and Integration with Tensor Parallelism for Spatial Scaling}

As discussed in the introduction section, Tensor Parallelism (TP) has emerged as the de facto standard for latency-oriented scaling in multi-GPU environments. We propose CTA-pipelining as an orthogonal spatial scaling technique that can be combined with TP. The evaluation is this section consists of two parts. First, we compare CTA-pipelining against TP on multi-layer GEMM. Second, we evaluates the integration of CTA-pipelining with TP, demonstrating how their combined application can further push the latency optimization frontier.

The baseline TP implementation mirrors the standard Megatron-LM paradigm for LLM MLP layers~\cite{shoeybi2019megatron}. Under standard TP for calculating $Y = XAB$, $A$ is sharded in a column-parallel fashion, and $B$ is sharded in a row-parallel fashion across all participating GPUs. An All-reduce collective operation is then required to sum the partial results. For the pure TP cases, cuBLAS kernels are again used to provide a strong baseline. The NCCL library is used for the All-reduce collective operations.

Note that splitting the input tensor $X$ while replicating the weight matrices $A$ and $B$ across devices to achieve Data Parallelism is not the standard practice for LLM deployments. In modern LLMs, transformer layers are repeated numerous times throughout the architecture, each containing multiple massive MLP weight matrices. As the total parameter size exceeds the memory capacity of a single GPU, replicating these weights across devices is impractical. Instead, weight matrices $A$ and $B$ are sharded across devices, while the input tensor $X$ can be replicated.

\subsubsection{Comparing CTA-Pipelining Against TP}






We perform our multi-layer GEMM comparison using the same experimental setup as the micro-batching evaluation above, where the number of GEMM layers increases with the GPU count. Under this setup, following standard Megatron-LM practices, TP requires an All-Reduce operation every two layers. We note that while a 2-layer structure is a more common pattern in real-world workloads, we include one example of multi-layer GEMM here for evaluation completeness. 

\begin{table}
    \centering
    \small
    \caption{Multi-layer GEMM (16384$\times$8192$\times$8192) latency reduction, comparing CTA-pipelining against Tensor Parallelism.}
    \label{tab:tp-speedup}
    \begin{tabular}{|c|c|c|c|c|}
        \hline
        Number of GPUs & 1 & 2 & 4 & 8 \\
        \hline
        Tensor Parallelism ($\mu$s) & 1080 & 1640 & 2636 & 4493 \\ 
        \hline
        CTA-Pipelining ($\mu$s) & 1080 & 1165 & 1418 & 1844 \\ 
        \hline
        Latency Reduction & - & 29.0\% & 46.2\% & 59.0\% \\ 
        \hline
    \end{tabular}
    \vspace{-3mm}
\end{table} 

The experiment results are shown in Table~\ref{tab:tp-speedup}. CTA-pipelining effectively reduces multi-layer GEMM latency by 29.0\%, 46.2\%, and 59.0\% on 2, 4, and 8 GPUs, respectively. The primary source of this benefit is that, within this setup, CTA-pipelining completely avoids All-Reduce communication. Conversely, TP requires frequent communication for every two layers. This demonstrates the advantage of using CTA-pipelining as a scaling method for suitable workflows that can execute in pure CTA-pipelining manner end-to-end.

As 2-layer GEMM represents a more realistic setup, similarly, we provide further evaluation across different input sequence length in Figure~\ref{fig:2-layer}. Result demonstrates the general benefit of CTA-pipelining against TP by saving communication time, despite the limitation on extremely small input sizes, which has been discussed in previous subsection. 

\begin{figure}[bp]
    \centering
    
    \begin{subfigure}[b]{0.9\columnwidth}
        \centering
        \includegraphics[width=\columnwidth]{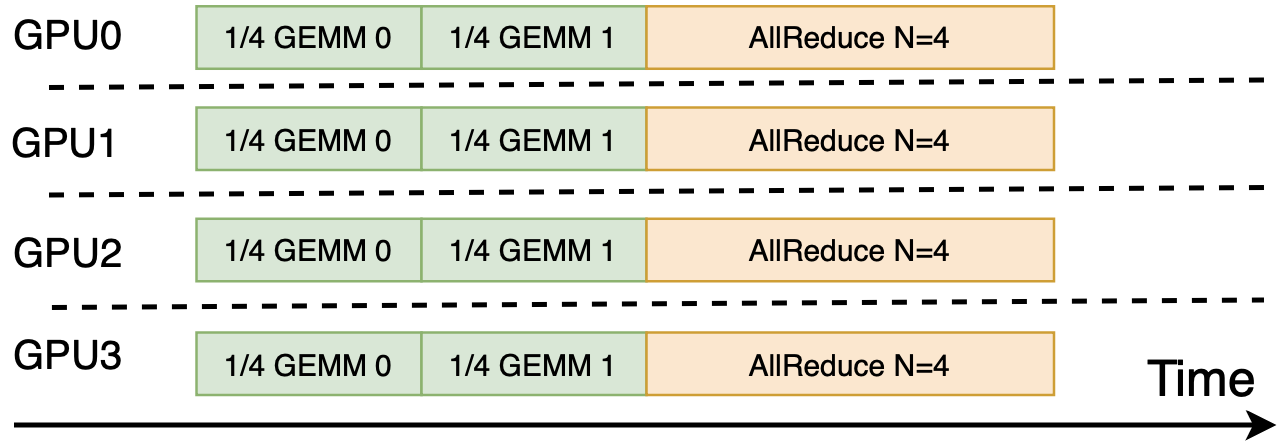}
        \caption{}
        \label{fig:tp-basic}
    \end{subfigure}

    \begin{subfigure}[b]{0.9\columnwidth}
        \centering
        \includegraphics[width=\columnwidth]{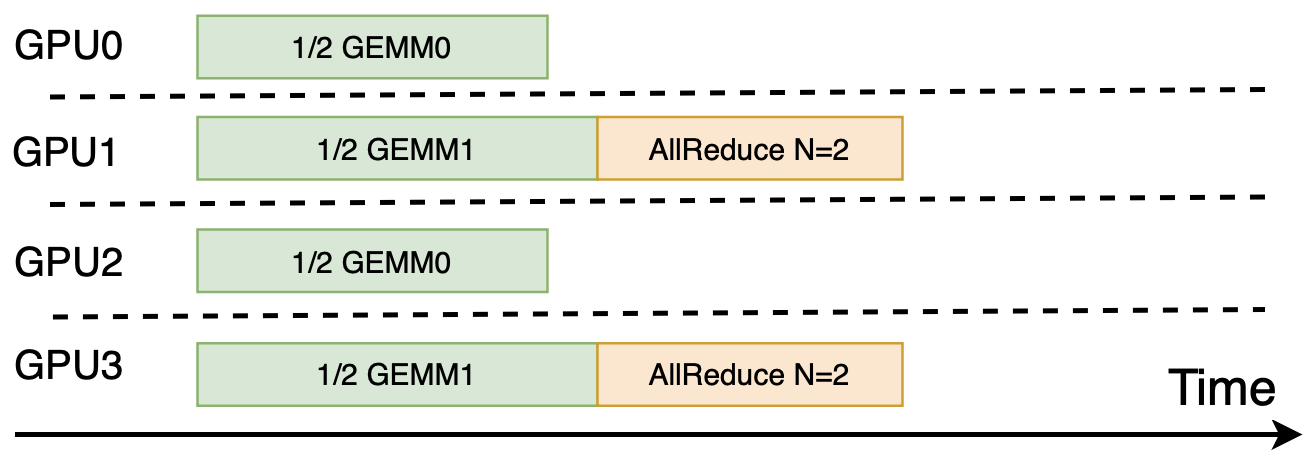}
        \caption{}
        \label{fig:tp-cta}
    \end{subfigure}
    
    \caption{The experiment setup for comparing executing 2 Layers of GEMM using (a) pure TP, and (b) the combination of TP and CTA-pipelining, using GPU=4 as an example.}
    \label{fig:tp-experiment}

    \vspace{-3mm}

\end{figure}

\begin{figure*}[htbp]
    \centering
    
    \begin{subfigure}[b]{0.32\textwidth}
        \centering
        \includegraphics[width=\textwidth]{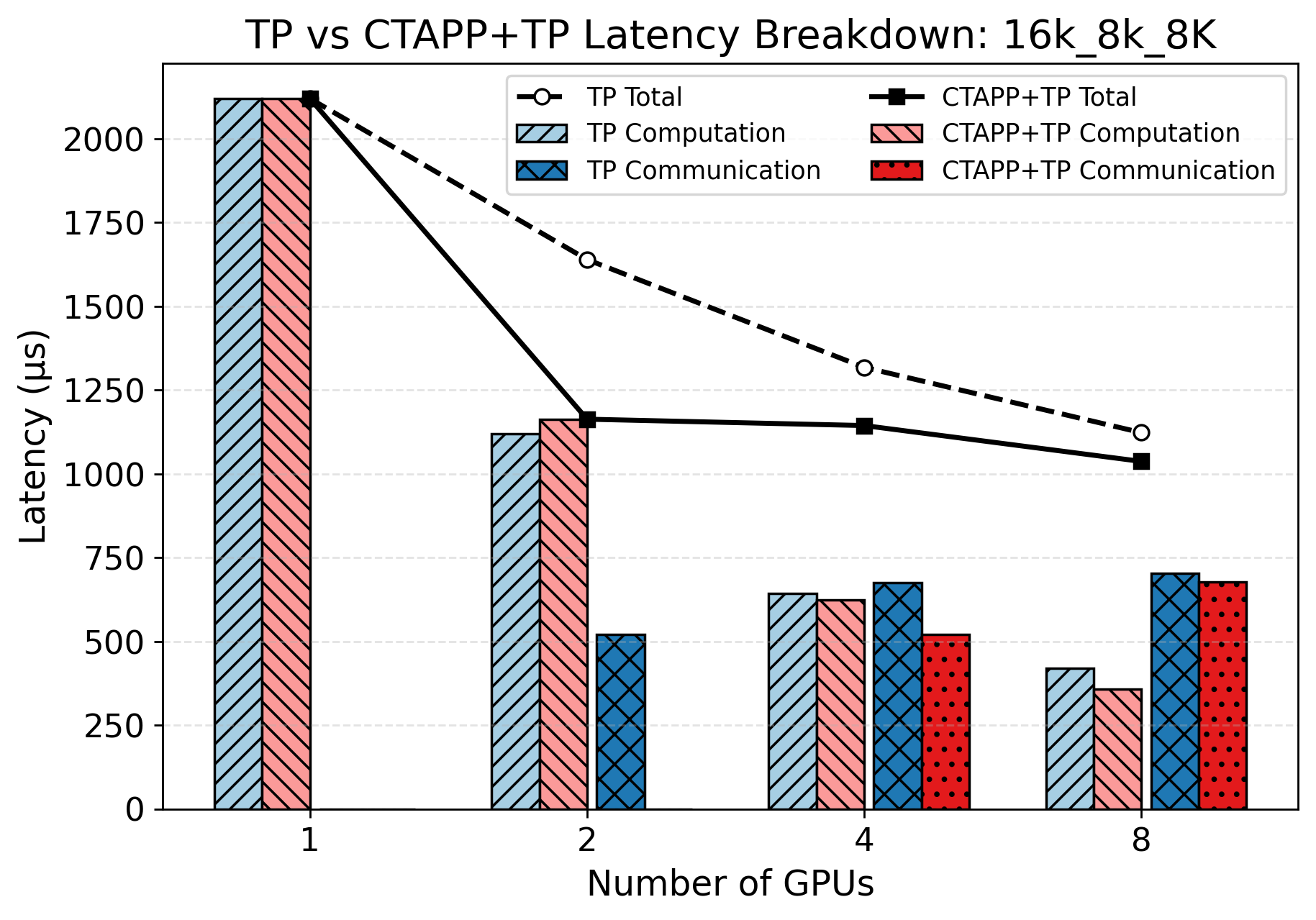}
        \caption{}
        \label{fig:tp-16k}
    \end{subfigure}
    \begin{subfigure}[b]{0.32\textwidth}
        \centering
        \includegraphics[width=\textwidth]{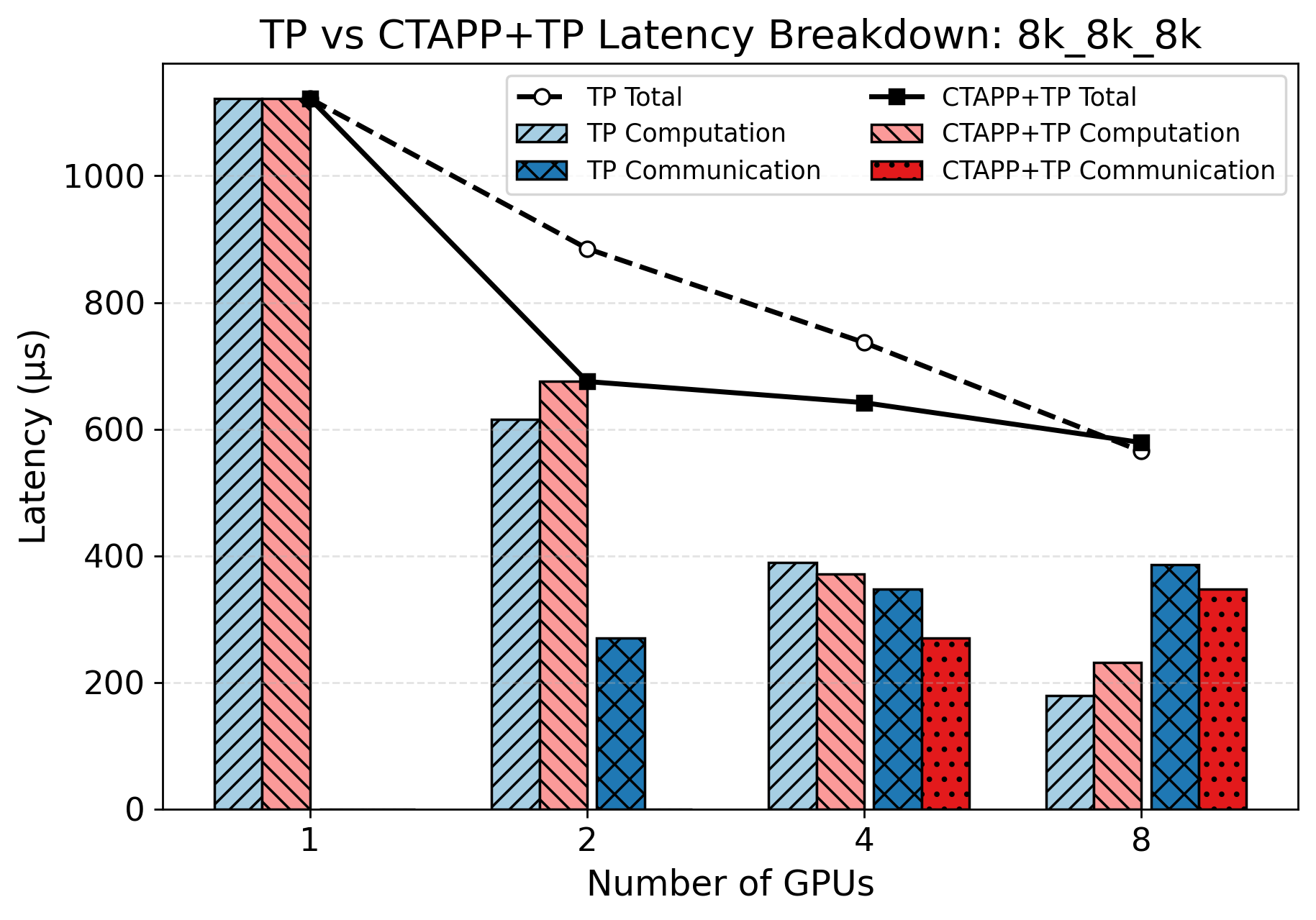}
        \caption{}
        \label{fig:tp-8k}
    \end{subfigure}
    \begin{subfigure}[b]{0.32\textwidth}
        \centering
        \includegraphics[width=\textwidth]{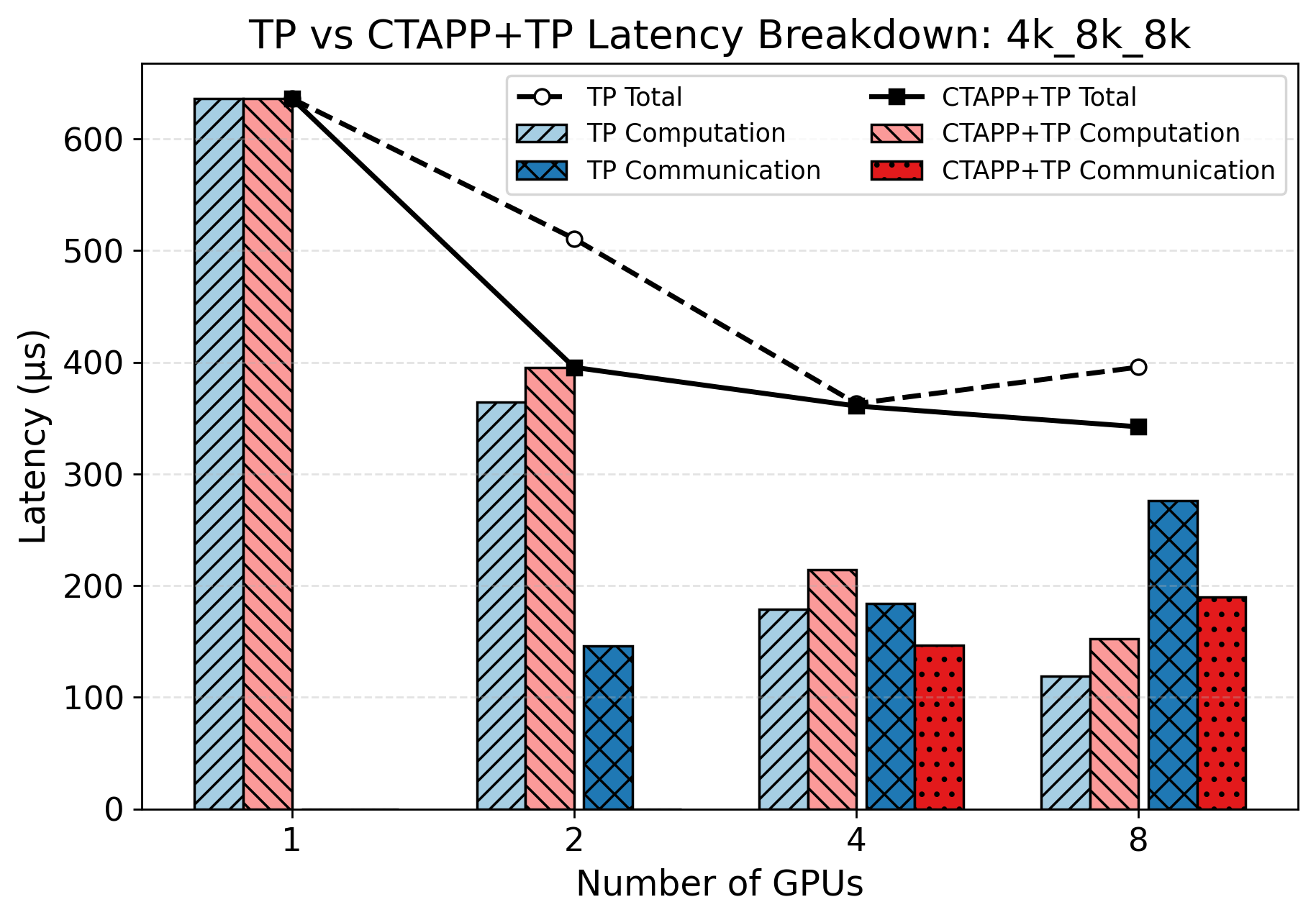}
        \caption{}
        \label{fig:tp-4k}
    \end{subfigure}

    \caption{The execution latency breakdown of 2-layer GEMM with different input sequence lengths, comparing pure TP and the combination of TP and CTA-pipelining.}
    \label{fig:tp-result}

    \vspace{-4mm}

\end{figure*}

\subsubsection{Combining CTA-pipelining with TP}

As demonstrated above, while CTA-pipelining provides significant benefits when workflows can execute entirely within this paradigm, some workflows lack enough kernels to form sufficient pipeline stages across multiple GPUs. In such scenarios, we purpose that, CTA-pipelining can be combined with TP as an orthogonal scaling dimension to further reduce latency. We demonstrate this capability using 2-layer GEMMs, representing the MLP layers.

To integrate CTA-pipelining with TP, we alter this sharding strategy. Rather than distributing $A$ and $B$ across all devices, we partition the hardware into 2-GPU groups. The weight matrices are then sharded evenly across these groups. Within any given group, the paired GPUs utilize CTA-pipelining to execute the local two-layer GEMM ($Y_i = XA_iB_i$). In this setup, the number of ranks involved in the final All-reduce operation is equal to the number of groups, halving the collective communication world size compared to the pure TP case. Figure~\ref{fig:tp-experiment} provides an example comparing pure TP with this combined CTA-pipelining and TP approach.

As described before, the weight matrices are fixed to be 8192 by 8192. We vary the input sequence length across 4096, 8192, and 16384 as representative user workload sizes. 

The results are shown in Figure~\ref{fig:tp-result}, with the lines showing the trend of total execution latency, and bars explicitly separate the computation and communication times for each setup. These results demonstrate that combining CTA-pipelining with TP provides a beneficial deployment strategy for multi-GPU scenarios, by pushing the latency frontier even further compared to pure TP deployment. 

The most direct benefit comes from the reduction in the world size involved in the All-reduce operation, which directly reduces the overall communication time. In some cases, such as in Figure~\ref{fig:tp-4k}, increasing pure TP degree can result in negative impact, since the communication time dominates, while combining with CTA-pipelining continues the latency reduction with more GPU resources. 

The other factor to consider is the pure computation time. When the TP degree is large, the input size per GPU becomes small, which can degrade kernel efficiency, similar to the micro-batching discussed earlier. By integrating CTA-pipelining, we effectively reduce the required TP degree, preserves larger matrix dimensions, and thereby benefits compute efficiency. However, executing via CTA-pipelining introduces additional pipeline's ramp-up delay (the one-wave CTA execution latency), which counteracts some of these computational gains. As a result, whether the computation time get benefit from combining with CTA-pipelining varies across different matrix sizes. 

In summary, the above experiments demonstrates that CTA-pipelining is a qualified spatial scaling method that is orthogonal to TP, allowing it to be seamlessly combined with TP for multi-GPU deployments. Compared to pure TP, this combined approach provides clear communication benefits, along with computational benefits in certain scenarios. Ultimately, this provides a powerful alternative for deploying multi-GPU workloads and holds the potential to further scale applications for lower latency, particularly when the whole workload can be executed in pure CTA-pipelining manner.

\section{Discussion} 

\subsection{Memory Consistency Using Lamport Synchronization}

As described in previous sections, to guarantee memory consistency during inter-kernel signaling, an explicit system-wide threadfence is issued after the output data write and before workqueue update. This ensures correct memory visibility ordering across the entire multi-GPU system. However, a system-wide threadfence is an expensive operation, especially when it must be issued for every single tile. Although we demonstrated that in warp-specialized multi-stage persistent kernels, this overhead can be hidden, classical kernels still suffer from this overhead. Furthermore, our empirical experiments show that even without an explicit threadfence, the probability of an out-of-order global write is low, motivating the exploration of further optimizations.

To address this, we evaluated an alternative method to guarantee memory consistency, referred to as Lamport synchronization. Under this approach, explicit threadfence are no longer used. Instead, the dependent data buffer between kernels is pre-initialized with dummy sentinel values (such as negative zero) that are unlikely to be produced during normal computation. Upon retrieving a ready CTA ID from the workqueue, the consumer kernel directly reads the dependent data from global memory, and performs an additional validation check. If the read data matches the dummy value, the kernel continuously re-reads from global memory until valid data is observed.

Despite its theoretical viability, Lamport synchronization introduces hardware and software implementation challenges. On Blackwell, data transfers directly to Tensor Memory (TMEM). As TMEM is only accessible to Tensor Cores, directly checking fetched data against dummy values is infeasible, while inferring correctness from final Tensor Core outputs instead dramatically increases programming complexity. Furthermore, Lamport synchronization might require tracking intermediate states (e.g., GEMM K-indices), which can increase shared memory pressure and degrades performance. Finally, the introduction of Lamport synchronization also breaks the simplicity of only adding prologue and epilogue code snippets to enable CTA-pipelining, creating challenge for automating the integration. 

As a result, we have excluded Lamport synchronization as the default configuration for our CTA-pipelining protocol. Nevertheless, we note that if Lamport synchronization were supported at the hardware level, where the memory subsystem natively checks sentinel values during global memory reads, this approach would become highly practical. Such hardware-assisted checking would benefit many applications far beyond CTA-pipelining.

\subsection{Tile-Based Models and Inter-Kernel Signaling}

It is natural to connect CTA-pipelining to the tile-based execution model, which has recently gained popularity with the introduction of frameworks like Triton~\cite{tillet2019triton} and cuTile~\cite{cutile}. If a GPU program is described using the tile abstraction, achieving tile-level inter-kernel pipeline execution becomes a highly intuitive process. The tile abstraction largely formalizes the input and output patterns of a kernel, thereby simplifying the dependency analysis between kernels. This demonstrates the potential of CTA-pipelining to serve as a widely adaptable technique for multi-kernel workflows in future GPU programming paradigms.

Building on these insights, enabling more efficient CTA-level inter-kernel signaling presents bigger potential impact. While our current CTA-pipelining protocol is fully functional on current hardware using strictly user-level CUDA code, it exposes a broader architectural paradigm shift. In modern GPUs, dense arithmetic is increasingly offloaded to specialized accelerators, frequently leaving the general-purpose Arithmetic Logic Units underutilized. Our methodology demonstrates how these idle general-purpose units can be effectively repurposed for inter-kernel orchestration. On the other hand, this also hints the potential for native driver-level support and future hardware integration for more efficient inter-kernel signalling.

\subsection{Communication Overlap and NVLink Topology}

The NVLink interconnect attempts to create the illusion of a unified multi-GPU system, where GPUs can access each other's memory as if it were local. To a large extent, this abstraction is successful, and our CTA-pipelining technique effectively utilize this illusion. However, current hardware still possesses physical limitations that prevent it from fully supporting this seamless abstraction. Analyzing how communication actually occurs during CTA-pipelining execution demonstrates these limitations, while also highlighting how more advanced hardware could eventually resolve them.

Within our 8-GPU B200 testbed, the devices are interconnected via NVLink and NVSwitch. While this fabric theoretically support a peer-to-peer bandwidth of 1.8 TB/s~\cite{nvhgx}, the routing is physically realized through the centralized switch. In contrast to traditional HPC topologies~\cite{hpctopology} that approximate non-blocking all-to-all connectivity, this star-like centralized architecture bottlenecks the maximum concurrent ingress and egress bandwidth per GPU.

Under the CTA-pipelining execution paradigm, this centralized NVLink topology restricts the full potential for compute-communication overlap. As the producer kernel writes output data directly to the consumer's device memory, leveraging the illusion of a shared-memory machine, it consumes NVLink bandwidth under the hood. Consequently, if the consumer device wants to achieve compute-communication overlap by communicating with a device other than the producer, its performance is still degraded by the producer's ongoing memory writes. Even though the producer is not actively involved in the consumer's secondary communication, both operations share the same physical interconnect links to the centralized NVLink Switch, resulting in contention.

Theoretically, these bottlenecks can be resolved by more advanced NVLink topologies, similar to the complex fabrics that currently exist for HPC clusters. For instance, the GB200 NVL72~\cite{nvidia2024gb200nvl72} system is capable to have more NVLink Switch. If we strategically place the workflow, such that a stack of kernels executing via CTA-pipelining is localized under a single NVLink Switch, while secondary communication is routed through a separate switch, compute-communication overlap becomes highly feasible. This would further increase the power of CTA-pipelining and hints at potential hardware-software co-design on upcoming large-scale multi-GPU shared-memory systems beyond NVL72.









\section{Conclusion}

In conclusion, this paper introduces CTA-pipelining, a latency-oriented spatial scaling paradigm that exploits modern multi-GPU shared-memory systems, aiming to reduce the single batch processing latency required by latest LLM inference workload. Through a prototype integration with NVIDIA CUTLASS, we demonstrate our protocol has minimal overhead, and can benefit from the warp-specialized multi-stage kernel design. When being used for spatial scaling, CTA-pipelining outperforms traditional micro-batch chunk pipelining and Tensor Parallelism on the example MLP setup, across various input sequence lengths. Furthermore, we establish CTA-pipelining as a general scaling dimension that composes orthogonally with Tensor Parallelism to further push the multi-GPU latency optimization frontier. As computation architectures evolves, CTA-pipelining offer a paradigm for future hardware-software co-design and next-generation multi-GPU execution model.


\bibliographystyle{IEEEtran}
\bibliography{sample}

\end{document}